\documentclass[a4paper,11pt]{article}
\usepackage{amsfonts}
\usepackage{amssymb,epic,eepic}
\usepackage{amsmath}
\author{Igor Bjelakovi\'c, Rainer Siegmund-Schultze\footnote{e-mail:\{igor,
siegmund\}@math.tu-berlin.de} \\
{\footnotesize Technische Universit\"at Berlin} \\
{ \footnotesize Fakult\"at II - Mathematik und Naturwissenschaften} \\
{ \footnotesize Institut f\"ur Mathematik MA 7-2} \\
{ \footnotesize Stra\ss e des 17. Juni 136} 
{ \footnotesize 10623 Berlin, Germany}}

\title{An Ergodic Theorem for the Quantum Relative Entropy}

\newcommand{\hr}{{\cal H}}

\newcommand{\cc}{{\mathbb C}}

\newcommand{\nn}{{\mathbb N}}
\newcommand{\idn}{\mathbf{1}}
\newcommand{\zz}{{\mathbb Z}}
\newcommand{\x}{{\mathbf x}}
\newcommand{\eps}{{\varepsilon}}        
\newcommand{\om}{\omega}
\newcommand{\kk}{{\mathbf k}}
\newcommand{\n}{{\mathbf n}}
\newcommand{\y}{{\mathbf y}}

\newtheorem{theorem}{Theorem}[section]         
\newtheorem{lemma}[theorem]{Lemma}             

\begin{document}

\maketitle
\begin{abstract}We prove the ergodic version of the quantum Stein's
  lemma which was conjectured by Hiai and Petz. The result provides an
  operational and statistical interpretation of the quantum relative
  entropy as a statistical measure of distinguishability, and contains as a
special case the quantum version of the Shannon-McMillan theorem for ergodic
states. A version of the quantum relative Asymptotic Equipartition Property (AEP)
is given.
\end{abstract}

\section{Introduction}
This paper can be seen as an extension of the article \cite{wir} by
Bjelakovi\'c, Kr\"uger, Siegmund-Schultze and Szko\l a, where instead of the von Neumann
mean entropy of an ergodic quantum state $\psi$ on a quasilocal algebra ${\mathcal A}^{\infty}$, dim ${\mathcal A}<\infty$, the
mean relative entropy $s(\psi,\varphi)$ of $\psi$ with respect to some stationary product state $\varphi$ is the basic quantity. If we choose
$\varphi$ to be the tracial state, the results here reduce to the quantum Shannon-McMillan theorem of \cite{wir}.
It turns out that the quantum mean relative entropy specifies the maximum exponential order, at which a typical subspace for
the ergodic state $\psi$ becomes untypical for the product state $\varphi$. Typical subspaces which asymptotically attain this
exponential order might be called \emph{maximally separating subspaces}. In particular, it turns out that projectors onto maximally separating
subspaces can be chosen in a way that an extended version of the quantum AEP is fulfilled: Each
one-dimensional projector dominated by the
maximally separating projector has an expected value with respect to $\psi$ which is of an exponential order given by the von Neumann
mean entropy of $\psi$ (in accordance with the quantum Shannon-McMillan theorem). The exponential order with respect to the reference
state $\varphi$ is larger exactly by the relative entropy. This is a complete analogue to the classical situation, which is included in the
result since ${\mathcal A}$ may be chosen abelian. 

We point to the fact, that already in the classical situation the i.i.d. assumption concerning the reference state cannot be weakened
very substantially, since there are examples with a reference process of very good mixing properties (B-process), but where a mean relative
entropy does not even exist, see \cite{shields}.

The methods used to derive the results are mainly based on the techniques developed in the paper Hiai/Petz \cite{petz} and in \cite{wir}.
The Hiai/Petz paper considered the class of completely ergodic states $\psi$, and only an (exact) upper bound for the separation order
was derived. Hiai and Petz were able to asymptotically reduce the problem to the abelian situation by constructing abelian sub-algebras 
with the remarkable property that they
simultaneously allow a restriction of the two states $\psi$, $\varphi$ asymptotically without distortion of their mean entropies and 
mean relative entropy.

With some extension of the methods used already in \cite{wir} it was possible to drop the assumption of complete ergodicity as well as to show
that the upper bound of the separation order is really a limit. This was conjectured already in Hiai/Petz\cite{petz} and proved for the 
case that both states are product (i.i.d.) states by Ogawa and Nagaoka in \cite{ogawa}.

\section{Asymptotics of the Quantum Relative Entropy}
In this section we shall state our main result.
We consider the lattice $\zz^{\nu}$. To each lattice site $\x\in \zz^{\nu}$ we 
assign a finite dimensional $C^{\ast}$-algebra ${\mathcal A}_{\x}$ being 
$\ast$-isomorphic to a fixed finite dimensional $C^{\ast}$-algebra ${\mathcal 
A}$. Recall that each finite dimensional $C^{\ast}$-algebra ${\mathcal A}$ can 
be thought of as $\bigoplus_{i=1}^{n}{\mathcal B}(\hr_{i})$ up to a
$\ast$-isomorphism where the $\hr_{i}$ are finite dimensional Hilbert spaces and ${\mathcal B}(\hr_{i})$ is the algebra of linear operators on $\hr_{i}$. For a
finite $\Lambda \subset \zz^{\nu}$ the algebra of local observables associated
to $\Lambda$ is defined by
\[{\mathcal A}_{\Lambda}:=\bigotimes_{\x\in\Lambda}{\mathcal A}_{\x}.\]
Clearly, this definition implies that for $\Lambda \subset \Lambda'$ we have 
${\mathcal A}_{\Lambda'}={\mathcal A}_{\Lambda}\otimes {\mathcal 
A}_{\Lambda'\setminus \Lambda}$. Hence there is a canonical embedding of 
${\mathcal A}_{\Lambda}$ into ${\mathcal A}_{\Lambda'}$ given by $a\mapsto 
a\otimes \idn_{\Lambda'\setminus \Lambda}$ for $a\in {\mathcal A}_{\Lambda}$ 
where $\idn_{\Lambda'\setminus \Lambda}$ denotes the identity of ${\mathcal 
A}_{\Lambda'\setminus \Lambda}$. The quasilocal $C^{\ast}$-algebra ${\mathcal
A}^{\infty}$ is the norm completion of the normed algebra $\bigcup_{\Lambda
\subset \zz^{\nu}}{\mathcal A}_{\Lambda}$ where the union is taken over all
finite subsets $\Lambda$. For a precise definition of the quasilocal algebra we 
refer to \cite{bratteli} and \cite{ruelle}. The group $\zz^{\nu}$ acts in a 
natural way on the quasilocal algebra ${\mathcal A}^{\infty}$ as translations by 
$\x \in \zz^{\nu}$. A translation $T(\x)$ by $\x$ associates to an $a\in 
{\mathcal A}_{\Lambda}$ the corresponding element $T(\x)a\in {\mathcal 
A}_{\Lambda+\x}$. This mapping $T(\x)$ extends canonically to a 
$\ast$-automorphism on ${\mathcal A}^{\infty}$ (cf. \cite{bratteli}, 
\cite{ruelle}). 

A state $\psi$ on ${\mathcal A}^{\infty}$ is a linear, positive and unital 
mapping of ${\mathcal A}^{\infty}$ into $\cc$. Each state $\psi$ on $ {\mathcal
A}^{\infty}$ corresponds to a compatible family $\{\psi^{(\Lambda)}\}_{\Lambda 
\subset \zz^{\nu}, \#\Lambda <\infty} $ where each $ \psi^{(\Lambda)}$ denotes 
the restriction of $\psi$ to the local algebra ${\mathcal A}_{\Lambda}$. The 
compatibility means that $\psi^{(\Lambda')}\!\upharpoonright {\mathcal 
A}_{\Lambda}=\psi^{(\Lambda)}$ for $\Lambda \subset \Lambda'$. A state $\psi$ is 
called stationary or translationally invariant if $\psi=\psi\circ T(\x)$ for all 
$\x\in \zz^{\nu}$. The convex set of stationary states is denoted by ${\mathcal 
T}({\mathcal A}^{\infty})$ and is compact in the weak$^{\ast}$-topology. The 
extremal points in ${\mathcal T}({\mathcal A}^{\infty})$ are the
\emph{ergodic} states. The basic theory of ergodic states is described in 
\cite{bratteli} and \cite{ruelle}.

In this article the local sets $\Lambda$ will mainly be boxes in $\zz^{\nu}$ which are
defined for $\n =(n_{1}, n_{2},\ldots ,n_{\nu}) \in \nn^{\nu}$ by
\[\Lambda(\n):=\{0,\ldots,n_{1}-1\}\times\ldots\times \{0,\ldots,n_{\nu}-1\},\]
and the cubic boxes of edge length $n \in \nn$ will be denoted by $\Lambda
(n)$.
The local algebras associated with $\Lambda (\n)$ will be denoted by ${\mathcal 
A}^{(\n)}$. In a similar fashion we will denote the restriction of a given state 
$\psi$ on ${\mathcal A}^{\infty}$ to the local algebra ${\mathcal A}^{(\n)}$ by 
$\psi^{(\n)}$. A state $\varphi$ is called a stationary product state if 
$\varphi^{(\n)}=(\varphi^{(1)})^{\otimes \Lambda(\n)}$ holds for each $\n \in \nn$.

The \emph{mean relative entropy} of a stationary state $\psi$ with respect to
the stationary product state $\varphi$ is defined by
\begin{eqnarray}\label{eq:def-mean-rel}
  s(\psi,\varphi):=\lim_{\Lambda (\n)\nearrow \nn^{\nu}}\frac{1}{\#\Lambda
(\n)}S(\psi^{(\n)},\varphi^{(\n)})=\sup_{\Lambda (\n)}\frac{1}{\#\Lambda 
(\n)}S(\psi^{(\n)},\varphi^{(\n)}).
\end{eqnarray}
The last equation is an easy consequence of the superadditivity of the relative
entropy of the state $\psi^{(\n)}$ with respect to the \emph{product} state 
$\varphi^{(\n)}$. For the convenience of the reader we recall the definition of 
the relative entropy of the state $\sigma$ on a finite dimensional $ 
C^{\ast}$-algebra ${\mathcal C}$ with respect to the state $\tau$:
\[ S(\sigma,\tau):=\left\{ \begin{array}{ll}  \textrm{tr}(D_{\sigma}(\log
D_{\sigma}-\log D_{\tau})) & \textrm{if  supp}(\sigma) \le \textrm{supp}(\tau)\\
\infty & \textrm{otherwise}
\end{array} \right.  \]
where $D_{\sigma}$ respectively $D_{\tau}$ denote the density operators of 
$\sigma$ respectively $\tau$ and supp$(\sigma)$ is the support projector of the
state $\sigma$. The important properties of the quantum relative entropy are 
summed up in the monograph \cite{ohya} by Ohya and Petz. Later on we will need 
the notion of the mean relative entropy of the stationary state $\psi$ with 
respect to the stationary product state $\varphi$ on the lattice
$G_{l}:=l\cdot\zz^{\nu}$, $l\ge 1$ an integer, given by
\begin{eqnarray}\label{eq:gl-rel-entr}
s(\psi,\varphi,G_{l}):= \lim_{\Lambda (\n)\nearrow \nn^{\nu}}\frac{1}{\#\Lambda 
(\n)}S(\psi^{(l\n)},\varphi^{(l\n)})=l^{\nu}s(\psi,\varphi). 
\end{eqnarray}

The basic quantity in this article is given by
\begin{eqnarray}\label{eq:beta-rel}
\beta_{\eps,\n}(\psi,\varphi):=\min\{\log\varphi^{(\n)}(q): q\in {\mathcal 
A}^{(\n)}\textrm{ projector and }\psi^{(\n)}(q)\ge 1-\eps\},
  \end{eqnarray}
where $\eps\in(0,1)$ and $\psi,\varphi\in {\mathcal T}({\mathcal A}^{\infty})$.
The quantities appearing in
(\ref{eq:beta-rel}) have a natural interpretation from the point of view of quantum statistical
hypothesis testing. Assume that under the hypothesis $H_{1}$ the state of the 
system is given by $\psi^{(\n)}$ and under the hypothesis $H_{2}$ the state is 
given by $\varphi^{(\n)}$. A projector $q\in{\mathcal A}^{(\n)}$ can be thought 
of as a decision rule: If a measurement of the projector $q$ on the system under 
consideration has outcome $1$, then the hypothesis $H_{1}$ is
accepted and the measurement outcome $0$ implies that $H_{2}$ is accepted. The 
quantities $\psi^{(\n)}(1-q)$ and $\varphi^{(\n)}(q)$ respectively describe the 
probabilities for the occurrence of the eigenvalue $0$ respectively $1$ if the 
hypothesis $H_{1}$ respectively $H_{2}$ was accepted, i.e. the error 
probabilities. The following theorem had been conjectured by Hiai and Petz in \cite{petz}.
\begin{theorem}\label{quant-stein}
Let $\psi$ be an ergodic state and let $\varphi$ be a stationary product state
on ${\mathcal A}^{\infty}$ with mean relative entropy $s(\psi,\varphi)<\infty$.
Then for all $\eps\in (0,1)$ we have
\[\lim_{\Lambda (\n)\nearrow \nn^{\nu}}\frac{1}{\#\Lambda(\n)}\beta_{\eps,\n}(\psi,\varphi)=-s(\psi,\varphi).\]
\end{theorem}
The fact that in the case where both states are stationary product states the limit
of $\frac{1}{\#\Lambda(\n)}\beta_{\varepsilon,\n}(\psi,\varphi)$ exists and equals $-s(\psi,\varphi)$,
had been shown by Ogawa and Nagaoka in \cite{ogawa}. Their proof requires some
deeper properties of so called quasi-entropies which are quantities defined
analogously to the relative entropy where an arbitrary  operator convex
function is used in their definition instead of  the function $-\log x$. We
shall derive now the existence of the limit in the general case. As an instrument
we will use some abelian approximation of all relevant quantities.

Projectors $q$ in (\ref{eq:beta-rel}) for which $\varphi^{(\n)}(q)$ is of the order $e^{-\#\Lambda(\n)s(\psi,\varphi)}$
and for which $\psi(q)\ge 1-\eps$ could be called
\emph{maximally separating projectors} and their range could be denoted as
\emph{maximally separating subspace}.
The proof of this theorem reveals some additional information about maximally separating projectors, which is collected
in the following
\begin{theorem}[quantum relative AEP]\label{qAEP}
Let \(\psi\) be an ergodic state with mean entropy $s(\psi)$ and let $\varphi$ be a stationary product state
on ${\mathcal A}^{\infty}$ with mean relative entropy $s(\psi,\varphi)<\infty$. Then
for all \( \varepsilon > 0\) there is an \( \mathbf{N}_{ \varepsilon} \in \nn^{\nu}\)
such that for all $\n \in \nn^{\nu}$ with \(\Lambda(\n) \supseteq \Lambda(\mathbf{N}_{ \varepsilon})\) there exists an
orthogonal projector \(p_{\n}(\varepsilon) \in {\cal A}^{(\n)}\) such that
\begin{enumerate}
\item
\(\psi^{(\n)}(p_{\n}(\varepsilon))\geq 1-\varepsilon\),
\item
for all minimal projectors \(p \in {\cal A}^{(\n)}\) with \(p \leq p_{\n}(\varepsilon)\) we have
\[e^{-\#(\Lambda(\n))(s(\psi)+ \varepsilon)} < \psi^{(\n)}(p) <
e^{-\#(\Lambda(\n))(s(\psi)- \varepsilon)},\]
and consequently
\( e^{\#(\Lambda(\n))(s(\psi)- \varepsilon)} < \textup{tr}_{\n}(p_{\n}(\varepsilon)) < e^{\#(\Lambda(\n))(s(\psi)+ \varepsilon)}.\)
\item
for all minimal projectors \(p \in {\cal A}^{(\n)}\) with \(p \leq p_{\n}(\varepsilon)\) we have
\[e^{-\#(\Lambda(\n))(s(\psi)+s(\psi,\varphi) + \varepsilon)} < \varphi^{(\n)}(p) <
e^{-\#(\Lambda(\n))(s(\psi)+s(\psi,\varphi) - \varepsilon)},\]
and consequently
\[e^{-\#(\Lambda(\n))(s(\psi,\varphi) + \varepsilon)} < \varphi^{(\n)}(p_{\n}(\varepsilon)) <
e^{-\#(\Lambda(\n))(s(\psi,\varphi) - \varepsilon)},\]
\end{enumerate}
\end{theorem}

In the case that the state $\varphi$ is the tracial state on ${\mathcal
A}^{\infty}$ Theorem \ref{qAEP} above is equivalent to the quantum
version of the Shannon-McMillan theorem proved in \cite{wir} (cf. Theorem
2.1 in \cite{wir}).

The proof of Theorem \ref{quant-stein} will make use of a classical law of large
numbers for the (classical) mean relative entropy.  
\begin{theorem}\label{relbrei-theo}
Let $A$ be a finite set and $P$ respectively $Q$ be an ergodic
respectively an i.i.d. probability measure on
$[A^{\zz^{\nu}},\mathfrak{A}^{\zz^{\nu}}]$, where
$\mathfrak{A}^{\zz^{\nu}} $ is the $\sigma$-algebra generated by the
cylinder sets. We have
\begin{eqnarray}\label{eq:asdiverg}
\lim_{\Lambda (\n)\nearrow \nn^{\nu}}\frac{1}{\#\Lambda 
(\n)}\log\frac{P^{(\n)}(\om_\n)}{Q^{(\n)}(\om_{\n})}
=D_{M}(P,Q)\qquad P-\textrm{almost surely},
\end{eqnarray}
where $D_{M}(P,Q)$ denotes the mean relative entropy of $P$ with respect to $Q$
and $\om_{\n}\in A^{\Lambda (\n)}$ are the components of $\om \in A^{\zz^{\nu}}$ corresponding to the box
$\Lambda (\n)$.
\end{theorem}
The proof of this classical assertion is an elementary application of the
Shannon-McMillan-Breiman theorem and the individual ergodic theorem. The higher
dimensional versions of these theorems needed in the present situation can be 
found in the article \cite{ornstein} by Ornstein and Weiss.

\section{Proof of the Main Theorems}
We start with the remark, that due to the assumption $s(\psi,\varphi)<\infty$ without any loss of generality we may assume
the state $\varphi^{(1)}$ to be faithful.
The following theorem and the Lemma \ref{asimpt} provide the tools to drop the condition of \emph{complete} ergodicity that was originally
used by Hiai and Petz to prove the assertion of Lemma \ref{lim_beta_rel<s}. Then we make use of the Hiai/Petz approximation Theorem \ref{hiai-petz-lemma} to construct two stochastic processes which
approximate the two states in the sense of entropy and relative entropy. The classical result Theorem \ref{relbrei-theo} immediately yields a
sequence of projectors which are separating with a rate given by the mean relative entropy. We still have to prove that this rate cannot be
beaten. This is done in Lemma \ref {lim_beta_rel>s}. The idea can be summarized as follows: Assume the existence of a better (higher rate) separating
subspace. It is easy to see, that projecting this subspace into the constructed typical subspace would yield another separating subspace which
would be 'better' than the constructed one. Now we make use of the classical relative AEP valid for the constructed subspace to exclude this possibility.
\begin{theorem}\label{ergod_components}
Let $\psi$ be an ergodic state on ${\mathcal A}^{\infty}$. Then for
every subgroup $G_{l}:=l \cdot \zz^{\nu}$, with $l>1$ an integer, there exists a 
${\kk}(l) \in {\nn}^{\nu}$ and a unique
convex  decomposition of $\psi $ into $G_{l}$-ergodic
 states $ \psi_{\x}$:

\begin{eqnarray}\label{ergodic_decomposition}
\psi =\frac{1}{\#\Lambda ({\kk}(l))}\sum_{\x\in \Lambda ({\kk}(l))} \psi_{\x}.
\end{eqnarray}
The $G_{l}$-ergodic decomposition (\ref{ergodic_decomposition}) has the 
following properties:
\begin{enumerate}
\item $k_{j}(l) \leq l $ and $k_{j}(l)|l $ for all \(j\in
\{1,\dots ,\nu\}\)

\item \(\{\psi_{\x}\}_{\x\in \Lambda ({\kk}(l))}= \{\psi_{0} \circ 
T(-\x)\}_{\x\in \Lambda
  ({\kk}(l))} \)

\item For every \(G_{l}\)-ergodic state \(\psi_{\x}\) in the
convex decomposition (\ref{ergodic_decomposition}) of \(\psi \) the mean entropy 
 with respect to \(G_{l}\), \(s(\psi_{\x}, G_{l})\), is
equal to the mean entropy
\(s(\psi, G_{l})\), i.e.
\begin{eqnarray}\label{equal_entropies}
s(\psi_{\x}, G_{l}) = s(\psi, G_{l}) 
\end{eqnarray}
for all $\x\in \Lambda ({\kk}(l))$.
\item For each \(G_{l}\)-ergodic state \(\psi_{\x}\) in the
convex decomposition (\ref{ergodic_decomposition}) of \(\psi \) and for every 
stationary product state $\varphi$ the mean relative entropy  with respect to 
\(G_{l}\), \(s(\psi_{\x},\varphi, G_{l})\), fulfills 
\begin{eqnarray}\label{equal_entropies}
s(\psi_{\x},\varphi, G_{l}) = s(\psi,\varphi, G_{l}) 
\end{eqnarray}
for all $\x\in \Lambda ({\kk}(l))$.
\end{enumerate}
\end{theorem}
\textbf{Proof of Theorem \ref{ergod_components}:} The first three items have 
been established in \cite{wir}, Theorem 3.1. \\
The proof of the last item is based on the monotonicity of the relative entropy 
and the usage of item 2. of the theorem. For each $\n\in \nn^{\nu}$ and
$\x\in \Lambda (\kk (l))$ we have $\psi_{\x}^{(\Lambda 
(l\n))}=\psi_{0}^{(\Lambda (l\n)-\x)}$ by the second item.
We consider the box $\tilde \Lambda (\n) $ containing $\Lambda (l\n)$ and each $\Lambda
(l\n)-\x$ defined by
\[ \tilde \Lambda (\n):=\{-l,\ldots ,ln_{1}-1\}\times \ldots \times
\{-l,\ldots ,ln_{\nu}-1\}, \]
and the box $\hat \Lambda (\n)$ contained in $\Lambda (l\n)$ and $\Lambda (l\n)-\x$
given by 
\[\hat \Lambda (\n):=\{0,\ldots ,l(n_{1}-1)-1\}\times \ldots \times \{0,\ldots
,l(n_{\nu}-1)-1\}.\] The volumes of these boxes are asymptotically equivalent in the sense that the quotient tends to one.
Hence using the observation above and twice the monotonicity of the relative 
entropy we obtain
\begin{eqnarray}
S(\psi_{0}^{(\tilde \Lambda (\n))},\varphi^{(\tilde \Lambda (\n))})&\ge&
S(\psi_{0}^{(\Lambda (l\n)-\x)},\varphi^{( \Lambda (l\n)-\x)})\nonumber \\
&=& S(\psi_{\x}^{(\Lambda (l\n))},\varphi^{( \Lambda (l\n)-\x)})\nonumber \\
&\ge & S(\psi_{\x}^{(\hat\Lambda (\n))},\varphi^{(\hat \Lambda (\n))}).\nonumber 
\end{eqnarray}
After dividing by $\#\Lambda (\n)$ and taking the limit $\Lambda (\n)\nearrow 
\nn^{\nu}$ this inequality chain shows that
\[ s(\psi_{0},\varphi,G_{l})\ge s(\psi_{\x},\varphi,G_{l})\]
holds. A similar argument using $\psi_{\x}^{(\Lambda
(l\n)+\x)}=\psi_{0}^{(\Lambda (l\n))}$
 shows that the reverse inequality is also valid. Hence we established
\[s(\psi_{0},\varphi,G_{l})= s(\psi_{\x},\varphi,G_{l}).\]
Since the mean relative entropy is affine in its first argument on the set of 
$G_{l}$-invariant states this implies 
\[ s( 
\psi_{\x},\varphi,G_{l})=s(\psi,\varphi,G_{l})=l^{\nu}s(\psi,\varphi).\qquad 
\Box\]

An important ingredient in our proof of the Theorem
\ref{quant-stein} is a result proved by Hiai and Petz in \cite{petz}. The 
starting point
is the spectral decomposition of the density operator $D_{\varphi^{(1)}}$
corresponding to the state $\varphi^{(1)}$ on ${\mathcal A}$:
\[ 
D_{\varphi^{(1)}}=\sum_{i=1}^{d}\lambda_{i}e_{i},\]
where $e_{i}$ are one-dimensional projectors. Clearly, the spectral 
representation of the 
 tensor product of $D_{\varphi^{(1)}}$ over a box $\Lambda (\y)$ can be written as
\[
D_{\varphi^{(\y)}}=\sum_{i_{1},\ldots,i_{N(\y)}=1}^{d}\lambda_{i_{1}}\cdots
\lambda_{i_{N(\y)}} e_{i_{1}}\otimes\ldots\otimes e_{i_{N(\y)}},
\] 
where we have chosen some enumeration $\{1,\ldots ,N(\y)\}$ of the points 
belonging to the box $\Lambda (\y)$. We have
$N(\y)=y_{1}\cdot\ldots\cdot y_{\nu}$.
Collecting all one dimensional projectors $e_{i_{1}}\otimes\ldots\otimes 
e_{i_{N(\y)}}$ which correspond to the same eigenvalue of $D_{\varphi^{(\y)}}$ we
can rewrite the last expression as
\begin{eqnarray}\label{eq:domultipl}
D_{\varphi^{(\y)}}=\sum_{ n_{1},\ldots n_{d}: n_{1}+\ldots
+n_{d}=N(\y)}\left(\prod_{k=1}^{d}\lambda_{k}^{n_{k}}\right)p_{n_{1},\ldots 
n_{d}},  
\end{eqnarray}
with 
\[ p_{n_{1},\ldots n_{d}}:=\sum_{(i_{1},\ldots, i_{N(\y)})\in I_{n_{1},\ldots 
n_{d}}}e_{i_{1}}\otimes\ldots\otimes e_{i_{N(\y)}},\]
 where
\[ I_{n_{1},\ldots n_{d}}:=\{(i_{1},\ldots, 
i_{N(\y)}):\#\{j:i_{j}=k\}=n_{k}\textrm{ for } 1\le k\le d\}.\]
We define the conditional expectation with respect to the trace by
\[ E_{\y}:{\mathcal A}^{(\y)}\to \bigoplus_{n_{1},\ldots ,n_{d}:\ n_{1}+\ldots
+n_{d}=N(\y)} p_{n_{1},\ldots ,n_{d}}{\mathcal A}^{(\y)}p_{n_{1},\ldots
,n_{d}},\]
\begin{eqnarray}\label{eq:tracecondexp}
 E_{\y}(a):=\sum_{n_{1},\ldots ,n_{d}:\ n_{1}+\ldots +n_{d}=N(\y)   
}p_{n_{1},\ldots ,n_{d}} a \ p_{n_{1},\ldots ,n_{d}}.
\end{eqnarray}
We are prepared to state the announced important result of Hiai and
Petz, Lemma 3.1 and Lemma 3.2 in \cite{petz}.
\begin{theorem}\label{hiai-petz-lemma}
If $\psi$ is a stationary state on ${\mathcal A}^{\infty}$ and ${\mathcal 
D}_{\y}$ is the abelian subalgebra of ${\mathcal A}^{(\y)}$ generated by
$\{p_{n_{1},\ldots n_{d}}D_{\psi^{(\y)}}p_{n_{1},\ldots n_{d}}\}_{n_{1},\ldots 
,n_{d}}\cup \{p_{n_{1},\ldots n_{d}}\}_{n_{1},\ldots ,n_{d}}$ then
\[
  S(\psi^{(\y)},\varphi^{(\y)})= S(\psi^{(\y)}\upharpoonright {\mathcal 
D}_{\y},\varphi^{(\y)}\upharpoonright {\mathcal D}_{\y}) +S(\psi^{(\y)}\circ
E_{\y} )-S(\psi^{(\y)}),
\]
and
\begin{eqnarray}\label{eq:entropyapprox}
  S(\psi^{(\y)}\circ E_{\y} )-S(\psi^{(\y)})\le d\log(\#\Lambda (\y)+1).
\end{eqnarray}
Consequently we have
\begin{eqnarray}\label{eq:abel-approx1}
\lim_{\Lambda (\y)\nearrow \nn^{\nu}}\frac{1}{\#\Lambda
(\y)}S(\psi^{(\y)}\upharpoonright {\mathcal
D}_{\y},\varphi^{(\y)}\upharpoonright {\mathcal D}_{\y})=s(\psi,\varphi).
\end{eqnarray}
\end{theorem}
\textbf{Remark:} This theorem had been proved by Hiai and Petz in \cite{petz}
for the one-dimensional
lattice. However, their proof extends canonically to the present situation.\\
Any abelian algebra ${\mathcal D}_{\y}$ in Theorem \ref{hiai-petz-lemma} can be
represented as
\begin{eqnarray}\label{eq:abelalg1}
{\mathcal D}_{\y}=\bigoplus_{i=1}^{a_{\y}} \cc \cdot f_{\y,i}
  \end{eqnarray}
where $\{f_{\y,i}\}_{i=1}^{a_{\y}}$ is the set of orthogonal minimal projectors
in
${\mathcal D}_{\y}$. For any $\y$ we introduce a maximally abelian refinement ${\mathcal B}_{\y}$ of ${\mathcal D}_{\y}$
by splitting each $f_{\y,i}$ into a sum of orthogonal and minimal (in the sense of the algebra ${\mathcal A}^{(\y)}$) projectors
$g_{\y,i,j}$ which leads to the representation
\begin{eqnarray}\label{eq:abelalg}
{\mathcal B}_{\y}=\bigoplus_{i=1}^{a_{\y}} \bigoplus_{j=1}^{b_{\y,i}}\cc \cdot g_{\y,i,j}.
\end{eqnarray}
By the monotonicity of the relative entropy we get
\begin{eqnarray}
S(\psi^{(\y)}\upharpoonright {\mathcal D}_{\y},\varphi^{(\y)}\upharpoonright {\mathcal D}_{\y}) \le
S(\psi^{(\y)}\upharpoonright {\mathcal B}_{\y},\varphi^{(\y)}\upharpoonright {\mathcal B}_{\y}) \le
S(\psi^{(\y)},\varphi^{(\y)}),
\end{eqnarray}
from which we deduce
\begin{eqnarray}\label{eq:abel-approx}
\lim_{\Lambda (\y)\nearrow \nn^{\nu}}\frac{1}{\#\Lambda
(\y)}S(\psi^{(\y)}\upharpoonright {\mathcal
B}_{\y},\varphi^{(\y)}\upharpoonright {\mathcal B}_{\y})=s(\psi,\varphi),
\end{eqnarray}
by (\ref {eq:abel-approx1}).
\\
We choose a
positive integer $ l $ and consider the decomposition of $\psi \in
{\partial}_{ ex } \mathcal{T} (\mathcal{ A^{ \infty } })$  
into states ${\psi}_{\x}$ being ergodic with respect to the
action of $G_{l}$, i.e. $\psi=\frac{1}{\#\Lambda (\kk (l))}\sum_{\x
  \in \Lambda (\kk (l))}{\psi}_{\x}$ in accordance with Theorem \ref{ergod_components}. Moreover, we consider a stationary product
state
$\varphi$. Note that this state is $G_{l}$-ergodic for each 
$l\in \zz$.
In order to keep our notation transparent we agree on following abbreviations: 
\[ s:=s(\psi ,\varphi ,{\zz}^{\nu})=s(\psi,\varphi),\]
i.e. the mean relative entropy of the state $\psi$ computed with respect to 
${\zz}^{\nu}$. We write  ${\mathcal B}_{l}$  for ${\mathcal B}_{(l,l,\ldots ,l)}$ and set
\[s_{\x}^{(l)}:=\frac{1}{\#\Lambda
(l)}S({\psi}_{\x}^{(\Lambda(l))}\upharpoonright {\mathcal B}_{l}
,{\varphi}^{(\Lambda(l))}\upharpoonright {\mathcal B}_{l}  )
\]
 and 
\[s^{(l)}:=\frac{1}{\#\Lambda
  (l)}S({\psi}^{(\Lambda (l))}\upharpoonright {\mathcal B}_{l} 
,{\varphi}^{(\Lambda(l))}\upharpoonright {\mathcal B}_{l}  ).\]
From the Theorem \ref{ergod_components} above we know that
\begin{equation} \label{eq:el}
s(\psi_{\x} ,\varphi ,G_{l})=s(\psi ,\varphi,G_{l})=l^{\nu} \cdot
s(\psi,\varphi),\qquad \forall
\x\in \Lambda (\kk (l)).
\end{equation}
For an arbitrarily chosen $\eta > 0$ let us define the set
\begin{equation}\label{eq:a}
 A_{l,\eta}:=\{\x \in \Lambda (\kk (l)): \
  s_{\x}^{(l)}< s-\eta\}.
\end{equation}
By \(A_{l,\eta}^{c}\) we denote its complement.
In the next lemma we shall show that the essential part  of $G_{l}$-ergodic
components of $\psi$ have the entropy per site close to $ s $  as $l$ becomes 
large, even if we restrict them to the abelian algebras ${\mathcal B}_{l}$.
\begin{lemma}\label{asimpt}
If $\psi$ is a ${\zz}^{\nu}$-ergodic state
and $\varphi$ a stationary product state on $ {\mathcal{A}}^{\infty}$ and if $
s(\psi,\varphi)<\infty$, then 
\[ \lim_{l \rightarrow \infty}\frac{\# A_{l,\eta}}{\# \Lambda (\kk (l))}=0\]
holds for every $\eta >0$.
\end{lemma}
\textbf{Proof of Lemma \ref{asimpt}:} We suppose, on the contrary, that there 
exists some $\eta_{0} >0$
such that $\limsup_{l}\frac{\# A_{l,\eta_{0}}}{\# \Lambda (\kk
  (l))}=a>0$. 
Then we can select a subsequence, which we denote again by $ (l)$ for 
simplicity, with the property  
\[ \lim_{l \rightarrow \infty}\frac{\#
  A_{l,\eta_{0}}}{\# \Lambda (\kk (l))}=a.\]
Using joint convexity of the relative entropy we have the following estimate:
\begin{eqnarray}\label{eq:densityentropy} 
\#\Lambda (\kk (l)) \cdot s^{(l)} & \le &
\sum_{\x \in \Lambda (\kk
  (l))}s_{\x}^{(l)}\nonumber \\
&=&\sum_{\x\in
 A_{l,\eta_{0}}}s_{\x}^{(l)}+\sum_{\x\in
 A_{l,\eta_{0}}^{c}}s_{\x}^{(l)} \nonumber \\
& \overset{(\ref{eq:a})} {<} & \# A_{l,\eta_{0}} \cdot (s-\eta_{0})+ \#
  A_{l,\eta_{0}}^{c} \cdot \max_{\x\in
  A_{l,\eta_{0}}^{c}}s_{\x}^{(l)}.
\end{eqnarray}
Employing that for the mean entropy
\[
s_{\x}^{(l)}\le \frac{1}{\#\Lambda (l)}S({\psi}_{\x}^{(\Lambda(l))} 
,{\varphi}^{(\Lambda(l))}  )\qquad \textrm{(by the monotonicity)}
\]
and 
\begin{eqnarray}s({\psi}_{\x},\varphi 
,G_{l})&=&\lim_{\Lambda(\mathbf{m})\nearrow \nn^{\nu}}\frac{1}{\#\Lambda
    (\mathbf{m})}S({\psi}_{\x}^{(l\mathbf{m})},\varphi^{(l\mathbf{m})})\nonumber
\\
&=&\sup_{\Lambda
    (\mathbf{m})}
\frac{1}{ \# \Lambda 
(\mathbf{m})}S({\psi}_{\x}^{(l\mathbf{m})},\varphi^{(l\mathbf{m})})\nonumber 
\end{eqnarray}
are fulfilled, we obtain an upper bound for the last term in (\ref{eq:densityentropy}):
\begin{eqnarray*}
\#
  A_{l,\eta_{0}}^{c} \cdot \max_{\x\in
  A_{l,\eta_{0}}^{c}}s_{\x}^{(l)}
&\le & \#
  A_{l,\eta_{0}}^{c} \cdot \max_{\x\in
  A_{l,\eta_{0}}^{c}}\frac{1}{l^{\nu}}s({\psi}_{\x},\varphi
,G_{l})
\\
& = & \#
  A_{l,\eta_{0}}^{c} \cdot s(\psi,\varphi) \qquad \textrm{
  (by(\ref{eq:el}))}.
\end{eqnarray*}
Inserting this in (\ref{eq:densityentropy}) and dividing by $\#\Lambda
(\kk (l))$ we obtain
\[s^{(l)}< \frac{\# A_{l,\eta_{0}}}{\# \Lambda (\kk (l))}(s-\eta_{0}) 
+\frac{\# A_{l,\eta_{0}}^{c}}{\# \Lambda (\kk (l))}s.\]
And after taking limits we arrive at the following contradictory inequality:
\[ s\le a(s-\eta_{0})+(1-a)s=s-a\eta_{0}<s,\]
since $\lim_{l\to \infty}s^{(l)}=s  $ by Theorem \ref{hiai-petz-lemma} and $ 
s<\infty$. Hence $a=0$.$\qquad \Box$ \newline
\begin{lemma}\label{lim_beta_rel<s}
Let $\psi$ be an ergodic state on $\mathcal{ A^{ \infty } }$ and let $\varphi$ 
be a stationary
product state on $\mathcal{ A^{ \infty } }$ fulfilling $ 
s(\psi,\varphi)<\infty$. Then
for every $\varepsilon \in (0,1)$
\[ \limsup_{\Lambda(\n) \nearrow \nn^{\nu}}\frac{1}{\#\Lambda(\n)} 
\beta_{\varepsilon,\n}(\psi,\varphi)\leq -s(\psi,\varphi).\]
\end{lemma}
\textbf{Proof of Lemma \ref{lim_beta_rel<s}:}
 We fix $\varepsilon >0$ and choose arbitrary $\eta$,
$\delta>0$. 
Consider the $G_{l}$-ergodic decomposition 
 \[\psi=\frac{1}{\#\Lambda (\kk(l))}\sum_{\x \in \Lambda (\kk(l))}\psi_{\x}\]
of $\psi$ for integers $l\geq1$. By Lemma \ref{asimpt} there is 
an integer $L\geq 1$ such that for any $l\geq L$ 
\[\frac{\varepsilon}{2} \geq \frac{1}{\# \Lambda (\kk(l))}\#{A_{l,\eta}}\geq 0\]
holds, where $A_{l,\eta}$ is defined by (\ref{eq:a}).
This inequality implies
\begin{equation} \label{eq:typical}
 \frac{1}{\#\Lambda (\kk(l))}\#{A_{l,\eta}^{c}} \cdot 
(1-\frac{\varepsilon}{2})\geq 1-\varepsilon. 
\end{equation}
Recall that by definition of $A_{l,\eta}$ we have
\begin{equation}\label{eq:goodentr}
\frac{1}{\#\Lambda(l)}S(\psi_{\x}^{(\Lambda(l))}\!\upharpoonright\mathcal{B}_{l},
\varphi^{(\Lambda (l))}\! \upharpoonright\mathcal{B}_{l})\geq 
s(\psi,\varphi)-\eta\qquad \textrm{for all } \x\in A_{l,\eta}^{c}.
\end{equation}
We fix an $l\geq L$
and consider the abelian quasi-local $C^{\ast}$-algebra
$\mathcal{B}_{l}^{\infty}$ built up
from $\mathcal{B}_{l}$. $\mathcal{B}_{l}^{\infty}$ is clearly a 
$C^{\ast}$-subalgebra of $\mathcal{A}^{\infty}$.
We set
\[m_{\x}:=\psi_{\x}\!\upharpoonright\mathcal{B}_{l}^{\infty} 
\textrm{ and } 
m_{\x}^{(\n)}:=\psi_{\x}\!\upharpoonright\mathcal{B}_{l}^{(\n)}.\]
Moreover we define
\[p:=\varphi \!\upharpoonright\mathcal{B}_{l}^{\infty}
\textrm{ and } p^{(\n)}:=\varphi \!\upharpoonright\mathcal{B}_{l}^{(\n)}.\]
The state $p$ is a $ G_{l}$-stationary product state. On the other hand,
Theorem 4.3.17. 
in \cite{bratteli} shows that the states $m_{\x}$ are $G_{l}$-ergodic. Due to 
the Gelfand isomorphism and the Riesz representation theorem we can (and shall) identify all
the states above with the probability measures on the corresponding maximal ideal
space of $ \mathcal{B}_{l}^{\infty}$. Since the algebra $\mathcal{B}_{l}$ is 
abelian and finite dimensional
this compact maximal ideal space can be thought of as $B_{l}^{\zz^{\nu}}$ for an
appropriately chosen finite set $B_{l}$. This is essentially the well-known 
Kolmogorov representation of a classical dynamical system.

By the definition of the measures $m_{\x}$ and $p$, monotonicity of the relative
entropy and the fourth item in Theorem \ref{ergod_components} we have
\begin{eqnarray} \label{dawarsmal}
D_{M}(m_{\x},p)\le s(\psi_{\x},\varphi,G_{l})=l^{\nu}s(\psi,\varphi)<\infty 
,
\end{eqnarray}
where $D_{M}(m_{\x},p)$ denotes the mean relative entropy of $m_{\x}$ with
respect to $p$.
Using the Theorem \ref{relbrei-theo} we see that
\begin{eqnarray}\label{eq:qrelbrei}
\lim_{\Lambda (\n)\nearrow \nn^{\nu}}\frac{1}{\Lambda 
(\n)}\log\frac{m_{\x}^{(\n)}(\omega_{\n})}{ 
p^{(\n)}(\omega_{\n})}=D_{M}(m_{\x},p)=:D_{M,\x}
\end{eqnarray}
$m_{\x}$-almost surely for all $\x\in \Lambda ({\kk}(l))$, where $\omega_{\n}\in
B_{l}^{\Lambda (\n)}$ are the components of $\omega\in B_{l}^{\zz^{\nu}}$ 
corresponding to the box $\Lambda (\n)$.
For each $\n$ and $\x\in A_{l,\eta}^{c}$ let
\begin{eqnarray}
 C_{\x}^{(\n)}& := & \left\{ \omega_{\n} \in B_{l}^{ (\n)}:\
|\frac{1}{\#\Lambda(\n)}\log\frac{
m_{\x}^{(\n)}(\omega_{\n})}{p^{(\n)}(\omega_{\n})} 
-D_{M,\x}|<\delta\right\}\nonumber \\
& = & \left\{ \omega_{\n} \in B_{l}^{ (\n)}|\ e^{\#\Lambda(\n) \cdot 
(D_{M,\x}-\delta)}<\frac{m_{\x}^{(\n)}(\omega_{\n})}{p^{(\n)}(\omega_{\n})}
<e^{\#\Lambda(\n) \cdot (D_{M,\x}+\delta)}\right\}.\nonumber 
\end{eqnarray}
The indicator function of $C_{\x}^{(\n)}$ corresponds to the projector 
$q_{\x}^{(\n)}\in \mathcal{B}_{l}^{(\n)}$. The definition of the set
$C_{\x}^{(\n)}$ implies a bound on the probability of this set with respect to 
the probability measure $p$:
\begin{eqnarray}\label{eq:p-prob}
 p^{(\n)}(C_{\x}^{(\n)})=\varphi (q_{\x}^{(\n)})&\leq &  
m_{\x}^{(\n)}(C_{\x}^{(\n)})e^{-\#\Lambda(\n) \cdot (D_{M,\x}-\delta)}\nonumber 
\\
& \leq & e^{-\#\Lambda(\n) \cdot (D_{M,\x}-\delta)}\nonumber \\
& \leq & e^{-\#\Lambda(\n) \cdot (\#\Lambda(l)(s(\psi,\varphi)-\eta)   
-\delta)},
\end{eqnarray}
because for all $\x \in A_{l,\eta}^{c}$ we have
\begin{eqnarray} \label{dmx}
D_{M,\x} &\overset{\textrm{by (\ref{eq:def-mean-rel})}}{\geq}&
D(m_{\x}^{(1)},p^{(1)})=S(\psi_{\x}^{(\Lambda(l))}\!\upharpoonright\mathcal{B}_{l},
\varphi^{(\Lambda (l))}\! \upharpoonright\mathcal{B}_{l}) \nonumber
\\
&\geq&
\#\Lambda(l)(s(\psi,\varphi)-\eta),
\end{eqnarray}
by (\ref{eq:goodentr}).
The limit assertion (\ref{eq:qrelbrei}) implies the existence of an $N\in \nn$
such that for
any $\n\in \nn^{\nu}$ with $\Lambda(\n)\supset \Lambda(N)$
\begin{equation} \label{eq:typ}
  m_{\x}^{(\n)}(C_{\x}^{(\n)})\geq 1-\frac{\varepsilon}{2},\qquad \forall
 \ \x\in A_{l,\eta}^{c}.
\end{equation}
For each $\y \in{\nn}^{\nu}$ with $y_{i}\geq Nl$ let
$y_{i}=n_{i}l+j_{i}$,
where $n_{i}\geq N$ and $0\leq j_{i}<l$. We set 
\[
q_{l\n}:=\bigvee_{\x\in A_{l,\eta}^{c}}q_{\x}^{(\n)} 
\] 
and denote by $q_{\y}$ the embedding of $q_{l\n}$ in $ \mathcal{A}^{(\y)}$.
By (\ref{eq:typ}) and (\ref{eq:typical}) we obtain
\begin{eqnarray}
\Psi^{(\y)}(q_{\y})&=&\frac{1}{\#\Lambda(k(l))}\sum_{\x\in
  \Lambda(k(l))} \psi_{\x}^{(\y)}(q_{\y})
 \nonumber \\
&\ge & \frac{1}{\#\Lambda(k(l))}\sum_{\x\in
  \Lambda(k(l))} \psi_{\x}^{(\y)}(q_{\x}^{(\n)})\nonumber \\
&\geq&\frac{1}{\#\Lambda(k(l))}\#
A_{l,\eta}^{c} \cdot (1-\frac{\varepsilon}{2})\geq 1-\varepsilon.\nonumber
\end{eqnarray}
Thus the condition in the definition of $\beta_{\varepsilon,\y}(\psi,\varphi)$ 
is satisfied.
We are now able to estimate $\beta_{\varepsilon,\y}(\psi,\varphi)$ using 
(\ref{eq:p-prob}):
\begin{eqnarray}
 \beta_{\varepsilon,\y}(\psi,\varphi) & \leq &
 \log
\varphi^{(\y)}(q_{\y})\nonumber \\
& \leq  & \log\sum_{\x\in
  A_{l,\eta}^{c}}e^{-\#\Lambda(\n) \cdot (D_{M,\x}-\delta)}  \nonumber \\
&\leq & \log(\# A_{l,\eta}^{c}\cdot e^{-\#\Lambda(\n) \cdot 
(\#\Lambda(l)(s(\psi,\varphi)-\eta)   -\delta)})\nonumber \\
&= & \log(\# A_{l,\eta}^{c}) -\# 
\Lambda(l\n)(s(\psi,\varphi)-\eta-\frac{\delta}{\#\Lambda(\l)}). \nonumber
\end{eqnarray}
This leads to
\[ \limsup_{\Lambda(\y) \nearrow \nn^{\nu}}\frac{1}{\#\Lambda(\y) 
}\beta_{\varepsilon,\y}(\psi,\varphi)\leq -s(\psi,\varphi)+\eta+\delta,\]
since $\# A_{l,\eta}^{c}$ does not depend on $\n$ and $\Lambda(\y)\nearrow 
{\nn}^{\nu}$ 
 if and only if $\Lambda(\n)\nearrow {\nn}^{\nu}$. Since $\eta$, $\delta>0$ were 
chosen arbitrarily we have
\[ \limsup_{\Lambda(\n) \nearrow \nn^{\nu}}\frac{1}{\#\Lambda(\n)}
\beta_{\varepsilon,\n}(\psi,\varphi)\leq -s(\psi,\varphi).\qquad \Box\]\\

Suppose we are given a sequence $(p_{\n})$ of projectors in ${\mathcal
A}^{(\n)}$ and a stationary state $\psi$ on ${\mathcal A}^{\infty}$ with mean 
entropy $s(\psi)$. We consider the positive operators
\begin{eqnarray}\label{eq:specrestriction}
p_{\n}D_{\psi^{(\n)}}p_{\n}=\sum_{i=1}^{d(\n)}\lambda_{\n,i}r_{\n,i}\qquad 
d(\n):=\mathrm{tr}(p_{\n}),
\end{eqnarray}
where the numbers $\lambda_{\n,i}$ are the eigenvalues and the $r_{\n,i}$ form a complete set of eigen-projectors of
$p_{\n}D_{\psi^{(\n)}}p_{\n}$. We set
\begin{eqnarray}\label{eq:goodboys}
T_{\n,\delta}:=\{i\in \{1,\ldots ,d(\n)\}:\ \lambda_{\n,i}\le e^{-\#\Lambda 
(\n)(s(\psi)-\delta)}\}
 \qquad \textrm{for }\delta>0,
\end{eqnarray}
and denote by $T_{\n,\delta}^{c}$ the complement of this set.
\begin{lemma}\label{goodprojectors}
Let $\psi$ be an ergodic state on ${\mathcal A}^{\infty}$ with mean entropy
$s(\psi)$ and let $(p_{\n})$ be a sequence of projectors in ${\mathcal
A}^{(\n)}$, respectively. If $p_{T_{\n,\delta}^{c}}$ is the projector corresponding to the set
$T_{\n,\delta}^{c}$ then
\[\lim_{\Lambda (\n)\nearrow \nn^{\nu}}\psi^{(\n)}(p_{T_{\n,\delta}^{c}})=0\]
for all $\delta >0$.
\end{lemma}
\textbf{Proof of Lemma \ref{goodprojectors}:}
We have
\[1\ge \sum_{i\in T_{\n,\delta}^{c}}\lambda_{\n,i}>e^{-\#\Lambda 
(\n)(s(\psi)-\delta)}\#T_{\n,\delta}^{c}=e^{-\#\Lambda 
(\n)(s(\psi)-\delta)}\mathrm{tr}(p_{T_{\n,\delta}^{c}}),\]
and, consequently
\[\frac{1}{\#\Lambda (\n)}\log (\mathrm{tr 
}(p_{T_{\n,\delta}^{c}}))<s(\psi)-\delta.\]
If we would have $\psi^{(\Lambda (\n))}(p_{T_{\n,\delta_{0}}^{c}})\ge a>0$ for
infinitely many $\n$ and some $a>0$ there would be a contradiction to
Proposition 2.1 in \cite{wir}, which implies that there is no sequence 
$(q_{\n})$ of projectors in ${\mathcal A}^{(\n)}$ with $\psi^{
(\n)}(q_{\n})\ge a>0$ and $\limsup_{\Lambda (\n)\nearrow
\nn^{\nu}}\frac{1}{\#\Lambda (\n)}\log (\mathrm{tr }(q_{\n}))<s(\psi)$.$\qquad 
\Box$
\begin{lemma}\label{lim_beta_rel>s}
Let $\psi$ be an ergodic state on $\mathcal{ A^{ \infty } }$ and let $\varphi$
be a stationary
product state on $\mathcal{ A^{ \infty } }$. Suppose that 
$s(\psi,\varphi)<\infty$ holds, then
for every $\varepsilon \in (0,1)$
\[ \liminf_{\Lambda(\n) \nearrow \nn^{\nu}}\frac{1}{\#\Lambda(\n)} 
\beta_{\varepsilon,\n}(\psi,\varphi)\geq -s(\psi,\varphi).\]
\end{lemma}
\textbf{Proof of Lemma \ref{lim_beta_rel>s}:} Let $(t_{\y})$ be a sequence of 
projectors, $t_{\y}\in{\mathcal A}^{(\y)}$, with $\Lambda (\y)\nearrow 
\nn^{\nu}$ and
\[
\liminf_{\Lambda (\y)\nearrow \nn^{\nu}}\frac{1}{\#\Lambda (\y)}\log 
\varphi^{(\y)}(t_{\y})<-s(\psi,\varphi).
 \]
Then there exists an $a>0$ and a subsequence, which we denote by $(t_{\y})$ for
notational simplicity, fulfilling
\begin{eqnarray}\label{eq:firstone}
\varphi^{(\y)}(t_{\y})<e^{-\#\Lambda (\y)(s(\psi,\varphi)+a)}.
\end{eqnarray}
We consider an integer $l\ge 1$ and the $G_{l}-$ergodic decomposition of $\psi$:
\[\psi=\frac{1}{\#\Lambda (\kk (l))}\sum_{\x\in \Lambda (\kk (l))}\psi_{\x}.\]
As in the proof of Lemma \ref{lim_beta_rel<s} for each $\eps, \eta >0$ we can choose $l$ in such a way
that we have
\begin{eqnarray}\label{eq:goodcomponents}
  \frac{\# A_{l,\eta}^{c}}{\#\Lambda (\kk (l))}(1-\frac{\eps}{2})\ge 1-\eps ,
\end{eqnarray}
where the set $A_{l,\eta}$ was defined by (\ref{eq:a}). Recall that we have by 
definition,
\[ \frac{1}{\#\Lambda (l)}S(\psi_{\x}^{(l)}\! \upharpoonright {\mathcal
B}_{l},\varphi^{(l)}\! \upharpoonright {\mathcal B}_{l})\ge 
s(\psi,\varphi)-\eta\quad \textrm{for all } \x\in A_{l,\eta}^{c}.\]
We consider again the abelian quasi-local algebra ${\mathcal B}_{l}^{\infty}$,
which will be identified with the algebra of continuous functions on the maximal
ideal space $B_{l}^{\infty}:=B_{l}^{\zz^{\nu}}$, bearing in mind that the restrictions of the
$G_{l}-$ergodic components of $\psi$ and $\varphi$ to this algebra are 
$G_{l}-$ergodic. We denote those restrictions by $m_{\x}$, $\x\in \Lambda (\kk 
(l))$, and $p$. As in the proof of Lemma \ref{lim_beta_rel<s} we can show that 
for $\delta >0$ the sets 
\begin{eqnarray}
  C_{\x ,\delta}^{(\n)}:=\{\omega_{\n}\in B_{l}^{(\n)}: \ e^{\#\Lambda (\n) 
(D_{M,\x}-\frac{\delta}{2})}< 
\frac{m_{\x}^{(\n)}(\omega_{\n})}{p^{(\n)}(\omega_{\n})}<e^{\#\Lambda (\n) 
(D_{M,\x}+\frac{\delta}{2})}\}\nonumber
\end{eqnarray}
fulfil
\begin{eqnarray}\label{eq:reltyp}
\lim_{\Lambda (\n)\nearrow \nn^{\nu}}m_{\x}^{(\n)}(C_{\x ,\delta}^{(\n)})=1  
\quad \textrm{for all } \x\in \Lambda(\kk(l)).
\end{eqnarray}
In a similar way, by employing the classical Shannon-McMillan theorem, we can 
see that for
\[
F_{\x ,\delta}^{(\n)}:= \{\omega_{\n}\in B_{l}^{(\n)}:e^{-\#\Lambda 
(\n)(h_{\x}+\frac{\delta}{2})}<m_{\x}^{(\n)}(\omega_{\n})<e^{-\#\Lambda 
(\n)(h_{\x}-\frac{\delta}{2})}\}, \]
we have
\begin{eqnarray}\label{eq:abstyp}
 \lim_{\Lambda (\n)\nearrow \nn^{\nu}}m_{\x}^{(\n)}(F_{\x ,\delta}^{(\n)})=1  
\quad \textrm{for all } \x\in \Lambda(\kk(l)),
\end{eqnarray}
where $h_{\x}$ denotes the Shannon entropy rate of $m_{\x}$. Each 
$\omega_{\n}\in  B_{l}^{(\n)}$ corresponds to a minimal projector $q_{\n}\in 
{\mathcal B}_{l}^{(\n)}\subset {\mathcal A}^{(l\n)}$. So, for all $\x\in A_{l,\eta}^{c}$ and for any
$\omega_{\n}\in C_{\x ,\delta}^{(\n)}\cap F_{\x ,\delta}^{(\n)}$ and 
corresponding $q_{\n}$ we have
\begin{eqnarray}\label{eq:estimate1}
 \varphi^{(l\n)}(q_{\n})&=& p^{(\n)}(\omega_{\n}) \nonumber \\
&>& e^{-\#\Lambda (\n)(D_{M,\x}+h_{\x}+\delta)} \nonumber \\
&\ge & e^{-\#\Lambda (\n)(s(\psi,\varphi)l^{\nu}+h_{\x}+\delta)}.
\end{eqnarray}
In fact, the first inequality follows from the definitions of the sets $C_{\x
,\delta}^{(\n)}$ and $ F_{\x ,\delta}^{(\n)}$ and the assumption that 
$\omega_{\n}\in C_{\x ,\delta}^{(\n)}\cap F_{\x ,\delta}^{(\n)}$. The second 
inequality is a consequence  of (\ref{dawarsmal}).
Moreover, by (\ref{dmx}) we get the upper estimate
\begin{eqnarray}
\varphi^{(l\n)}(q_{\n})&=& p^{(\n)}(\omega_{\n}) \nonumber \\
&<& e^{-\#\Lambda (\n)(D_{M,\x}+h_{\x}-\delta)} \nonumber \\
&\le & e^{-\#\Lambda (\n)((s(\psi,\varphi)-\eta)l^{\nu}+h_{\x}-\delta)} \label{dmx1}.
\end{eqnarray}
Next, observe that the representation (\ref{eq:abelalg}) of ${\mathcal
B}_{l}$ implies that
\[D_{\psi_{\x}^{(l)}\!\upharpoonright {\mathcal
B}_{l}}=\sum q_{l,i}D_{\psi_{\x}^{(l)}}q_{l,i},  \]
where $(q_{l,i})$ is a complete set of minimal eigen-projectors of
$D_{\varphi^{(l)}}$.
We get
\begin{eqnarray}\label{eq:mischterm}
\textrm{tr}(D_{\psi_{\x}^{(l)}\!\upharpoonright {\mathcal B}_{l}}\log D_{\varphi^{(l)}}
)&=&\textrm{tr}(\sum q_{l,i}D_{\psi_{\x}^{(l)}}q_{l,i}\log D_{\varphi^{(l)}})\nonumber \\
&=& \textrm{tr}(D_{\psi_{\x}^{(l)}}\sum q_{l,i}(\log D_{\varphi^{(l)}}
)q_{l,i}   )\nonumber \\
&=& \textrm{tr}(D_{\psi_{\x}^{(l)}}\log D_{\varphi^{(l)}}).
\end{eqnarray}
Finally, by our assumption that $s(\psi,\varphi)<\infty$ holds we
know that
$s(\psi,\varphi)=-s(\psi)-\textrm{tr}(D_{\psi^{(1)}}\log D_{\varphi^{(1)}})$. 
Using the product structure of the state $\varphi$ and the fact just derived we 
obtain
\begin{eqnarray}\label{eq:first}
S(\psi_{\x}^{(l\n)}\! \upharpoonright
{\mathcal B}_{l}^{(\n)},\varphi^{(l\n)}\! \upharpoonright {\mathcal
B}_{l}^{(\n)})&=&-H(m_{\x}^{(\n)})-\textrm{tr}(D_{\psi_{\x}^{(l\n)}\!\upharpoonright {\mathcal
B}_{l}^{\n}}\log D_{\varphi^{(l\n)}})\nonumber \\
&=& -H(m_{\x}^{(\n)})-\#\Lambda
(\n)\textrm{tr}(D_{\psi_{\x}^{(l)}\!\upharpoonright {\mathcal B}_{l}}\log 
D_{\varphi^{(l)}})\nonumber \\ 
&\overset {\textrm{(by (\ref{eq:mischterm}))}}{=}& -H(m_{\x}^{(\n)})-\#\Lambda (\n)(\textrm{tr}(D_{\psi_{\x}^{(l)}}\log
D_{\varphi^{(l)}}). \nonumber \\
\end{eqnarray}
On the other hand we have for all $\x \in A_{l,\eta}^{c}$ 
\begin{eqnarray}\label{eq:second}
S(\psi_{\x}^{(l\n)}\! \upharpoonright {\mathcal B}_{l}^{(\n)},\varphi^{(l\n)}\! 
\upharpoonright {\mathcal B}_{l}^{(\n)})&\ge & \#\Lambda (\n)( 
-H(m_{\x}^{(1)})-\textrm{tr}(D_{\psi_{\x}^{(l)}\!\upharpoonright {\mathcal 
B}_{l}}\log D_{\varphi^{(l)}}))\nonumber \\
& & \textrm{(by the subadditivity of the entropy)}\nonumber \\
&=&\#\Lambda (\n)S(\psi_{\x}^{(l)}\!\upharpoonright {\mathcal 
B}_{l},\varphi^{(l)}\!\upharpoonright {\mathcal B}_{l})\nonumber \\ 
&\ge & \#\Lambda (\n)l^{\nu}(s(\psi,\varphi)-\eta) \nonumber \\
& & \textrm{(by the choice of algebra } {\mathcal B}_{l})\nonumber \\
&=& \#\Lambda (\n)l^{\nu}(-s(\psi)-\textrm{tr}(D_{\psi^{(1)}}\log
D_{\varphi^{(1)}})-\eta).\nonumber\\
\end{eqnarray}
The equation chain (\ref{eq:first}) and the inequality chain (\ref{eq:second}) 
imply
that
\begin{eqnarray}\label{eq:entropyest} 
-H(m_{\x}^{(\n)})&-&\#\Lambda (\n)(\textrm{tr}(D_{\psi_{\x}^{(l)}}\log 
D_{\varphi^{(l)}}))\nonumber \\
&\ge&  \#\Lambda (\n)l^{\nu}(-s(\psi)-\textrm{tr}(D_{\psi^{(1)}}\log 
D_{\varphi^{(1)}})-\eta). 
\end{eqnarray}
Note that the third and the fourth item of Theorem \ref{ergod_components} show
that
\[ \textrm{tr}(D_{\psi_{\x}^{(l)}}\log D_{\varphi^{(l)}})=l^{\nu} 
\textrm{tr}(D_{\psi^{(1)}}\log D_{\varphi^{(1)}}),\]
in the case where $s(\psi,\varphi)<\infty$.
Dividing (\ref{eq:entropyest}) by $-\#\Lambda (\n)$ and taking the limit
$\Lambda (\n)\nearrow \nn^{\nu}$ leads to
\begin{eqnarray}\label{estimate:hx}
l^{\nu}s(\psi)\le h_{\x}\le l^{\nu}(s(\psi)+\eta),
\end{eqnarray}
where the lower bound simply follows from the fact that the entropy on a maximally abelian subalgebra is
not less than the entropy on the algebra.
This inequality, (\ref{eq:estimate1}) and (\ref{dmx1})
imply:
\begin{eqnarray}\label{eq:estimate}
e^{-\#\Lambda (l\n)(s(\psi,\varphi)+s(\psi)+\eta+\delta/l^{\nu})}\le \varphi^{(l\n)}(q_{\n})
\le e^{-\#\Lambda (l\n)(s(\psi,\varphi)+s(\psi)-\eta-\delta/l^{\nu})}.
\end{eqnarray}

Let $p_{l\n}$ denote the projector corresponding to the set
${\bigcup}_{\x \in A_{l,\eta}^{c}}C_{\x,\delta}^{(\n)}\cap F_{\x,\delta}^{(\n)}$ and $p_{\n,\x}$ be the
projector which corresponds to $C_{\x,\delta}^{(\n)}\cap F_{\x,\delta}^{(\n)}$.
For sufficiently large $\n \in \nn^{\nu}$ we have
\[\psi_{\x}(p_{\n,\x})\ge 1-\frac{\eps}{2}\qquad \textrm{by (\ref{eq:reltyp})
and (\ref{eq:abstyp})},\]
and hence by (\ref{eq:goodcomponents})
\begin{eqnarray}\label{eq:important}
\psi ( p_{l\n})&\ge& \frac{1}{\#\Lambda (\kk (l))}\sum_{\x \in 
A_{l,\eta}^{c}}\psi_{\x}(p_{\n,\x})\nonumber \\
&\ge & \frac{1}{\#\Lambda (\kk (l))}\#A_{l,\eta}^{c}(1-\frac{\eps}{2})\ge 
1-\eps.
\end{eqnarray}
Any $t_{\y}$ fulfilling (\ref{eq:firstone}) can be embedded in an appropriately
chosen ${\mathcal A}^{(l\n)}$. Indeed, choose the unique $\n \in \nn^{\nu}$ such 
that $(n_{i}-1)l\le y_{i}<n_{i}l$ for all $i\in \{1,\ldots ,\nu\}$ and set
\[e_{l\n}:=t_{\y}\otimes \idn_{\Lambda (l\n)\setminus \Lambda (\y)}.\] 
Then we have 
\begin{eqnarray}\label{eq:equalrestrict}
\psi^{(l\n)}(e_{l\n})=\psi^{(\y)}(t_{\y}),
\end{eqnarray}
and 
\begin{eqnarray}
\psi^{(l\n)}(e_{l\n})&=&\psi^{(l\n)}(e_{l\n}p_{l\n})+ \psi^{(l\n)}(e_{l\n}(\idn 
-p_{l\n}))\nonumber \\ 
&\le & \psi^{(l\n)}(e_{l\n}p_{l\n})+ \eps \qquad \textrm{(by 
(\ref{eq:important}))}.\nonumber
\end{eqnarray}
Applying this argument once more we obtain
\begin{eqnarray}\label{eq:sepsubspace}
 \psi^{(l\n)}(e_{l\n}) &\le &\psi^{(l\n)}(p_{l\n}  e_{l\n}p_{l\n})+ 2\eps
\nonumber\\
&=& \textrm{tr}(p_{l\n}D_{\psi^{(l\n)}}p_{l\n}e_{l\n})+2\eps. 
\end{eqnarray}
In the final step we will prove that the first term in the last line above can 
be made arbitrarily small.
Using the notation from (\ref{eq:specrestriction}), (\ref{eq:goodboys}) and 
applying Lemma
\ref{goodprojectors} we know that
\begin{eqnarray}\label{eq:reduction}
\textrm{tr}(p_{l\n}D_{\psi^{(l\n)}}p_{l\n}e_{l\n})&\le &  e^{-\#\Lambda
(l\n)(s(\psi)-\delta)}\sum_{i\in 
T_{l\n,\delta}}\textrm{tr}(r_{l\n,i}e_{l\n})+\eps\nonumber \\
&\le &  e^{-\#\Lambda 
(l\n)(s(\psi)-\delta)}\sum_{i=1}^{\textrm{tr}(p_{l\n})}\textrm{tr}(r_{l\n,i}e_{l\n})+\eps
\nonumber \\
&=& e^{-\#\Lambda (l\n)(s(\psi)-\delta)}\textrm{tr}(p_{l\n}e_{l\n})+\eps,
\end{eqnarray}
for sufficiently large $\n \in \nn^{\nu}$. We represent the projector $p_{l\n}$ 
as a sum of unique minimal projections in ${\mathcal B}_{l}^{(\n)}$:
\[p_{l\n}=\sum_{i=1}^{\textrm{tr}(p_{l\n})}q_{\n,i}.\]
Hence by (\ref{eq:estimate}) we have
\begin{eqnarray}\label{eq:final}
\textrm{tr}(p_{l\n}e_{l\n}) &\le&   e^{\#\Lambda
(l\n)(s(\psi,\varphi)+s(\psi)+\eta+\delta/l^{\nu})}\sum_{i=1}^{\textrm{tr}(p_{l\n})}\varphi^{(l\n)}(q_{\n,i})\textrm{tr}(q_{\n,i}e_{l\n})\nonumber \\
&=& e^{\#\Lambda
(l\n)(s(\psi,\varphi)+s(\psi)+\eta+\delta/l^{\nu})}\varphi^{(l\n)}(p_{l\n}e_{l\n})\nonumber \\
&\le&e^{\#\Lambda
(l\n)(s(\psi,\varphi)+s(\psi)+\eta+\delta/l^{\nu})}\varphi^{(\y)}(t_{\y})\nonumber \\
&\le & e^{\#\Lambda
(l\n)(s(\psi,\varphi)+s(\psi)+\eta+\delta/l^{\nu})}e^{-\#\Lambda
(\y)(s(\psi,\varphi)+a)}\textrm{ (by (\ref{eq:firstone}))},\nonumber \\
\end{eqnarray}
since the minimal projectors $ q_{\n,i}$ correspond to the eigen-vectors
of $D_{\varphi^{(l\n)}}$. Inserting this into (\ref{eq:reduction}) we obtain
\begin{eqnarray}
\textrm{tr}(p_{l\n}D_{\psi^{(l\n)}}p_{l\n}e_{l\n})&\le & e^{-\#\Lambda 
(l\n)((\frac{\#\Lambda (\y)}{\#\Lambda (l\n)}-1)s(\psi,\varphi)+\frac{\#\Lambda 
(\y)}{\#\Lambda (l\n)}a  -\delta-\eta -\frac{\delta}{l^{\nu}})}+\eps. \nonumber 
\end{eqnarray}
Note that $\lim_{\Lambda (\y)\nearrow \nn^{\nu}} \frac{\#\Lambda (\y)}{\#\Lambda 
(l\n)}=1$, and that $a>0$. Hence if we choose $\delta , \eta $ small enough and 
$\n$ large enough we can achieve that the exponent in the last inequality 
becomes negative eventually. This inequality toghether with
(\ref{eq:sepsubspace}) and (\ref{eq:equalrestrict}) shows that
\[\lim_{\Lambda (\y)\nearrow \nn^{\nu}}\psi(t_{\y})=0.\qquad \Box \]

The combination of Lemma \ref{lim_beta_rel<s} and Lemma \ref{lim_beta_rel>s} yields the assertion
of Theorem \ref{quant-stein}.
To derive Theorem \ref{qAEP}, consider the projectors $p_{l\n}$ constructed in the proof of
Lemma \ref{lim_beta_rel>s}. They can be written as the sum of minimal projectors $q_{\n}$ fulfilling
(\ref{eq:estimate}). From the fact, that these are minimal eigen-projectors for $D_{\varphi^{(\n)}}$ we derive
that (\ref{eq:estimate}) is valid even if we replace $q_{\n}$ by any minimal projector which is dominated
by $p_{l\n}$. So we see that, if we would choose the projectors $p_{\n}(\varepsilon)$ to be specified for
Theorem \ref{qAEP} just as $p_{l\n}$ for boxes with edge lengths being multiples of $l$, item 3 would
be satisfied for large $\n$, supposed $l$ is (fixed but) large enough. Item 1 is fulfilled for large $\n$
in view of (\ref{eq:important}). Observe now that embedding the projectors $p_{l\n}$ into $\mathcal{A}^{(\y)}$
for $0 \le \y-l\n<(l,l,\ldots,l)$ leads us to a family of projectors $(p_\n)$ still fulfilling item 1. Item 3
is satisfied by this family, too, since $\varphi^{(1)}$ was supposed faithful. In order to ensure that item 2
is fulfilled, the constructed family of projectors $(p_{\n})$ has to be modified. Represent $p_{\n}$ as a sum
of eigen-projectors of the operator $p_{\n}D_{\psi^{(\n)}}p_{\n}$. Now from the definition of the sets
$F^{(\n)}_{\x,\delta}$ we easily conclude that $\textrm{tr}(p_{\n})<e^{\#\Lambda(\n)(s(\psi)+\varepsilon)}$
can be guaranteed for large $\n$. So the asymptotic contribution to $\psi^{(\n)}(p_{\n})$ of eigen-values of
$p_{\n}D_{\psi^{(\n)}}p_{\n}$ of magnitude exponentially smaller than
$e^{-\#\Lambda(\n)(s(\psi)+\varepsilon)}$ can be neglected. The asymptotic contribution to
$\psi^{(\n)}(p_{\n})$ of eigen-values of
$p_{\n}D_{\psi^{(\n)}}p_{\n}$ of magnitude exponentially larger than
$e^{-\#\Lambda(\n)(s(\psi)-\varepsilon)}$ can be neglected, too, because of Lemma \ref{goodprojectors}.
So we may omit the eigen-projectors corresponding to either too large or too small eigen-values from the sum,
getting a modified family $(p'_{\n})$, which additionally fulfils item 2. This proves Theorem \ref {qAEP}.
\\
\newline
\textbf{Acknowledgements} We are very thankful to our dear colleagues Tyll Kr\"uger, Ruedi Seiler and
 Arleta Szko\l a for their constant interest and encouragement during the preparation of this
article and for many very helpful comments and improvements.

This work was supported by the DFG via SFB 288 'Differentialgeometrie und Quantenphysik' at TU Berlin.


\begin{thebibliography}{99}

\bibitem{wir}
I. Bjelakovi\'c, T. Kr\"uger, Ra. Siegmund-Schultze, A. Szko\l a, The 
Shannon-McMillan Theorem
for Ergodic Quantum Lattice Systems, arXiv.org, math.DS/0207121, to
appear in \emph{Inventiones Mathematicae} (2003)


\bibitem{bratteli}
O.Bratteli, D.W. Robinson, Operator Algebras and Quantum Statistical
Mechanics I, Springer, New York 1979


\bibitem{petz} 
F. Hiai, D. Petz, The Proper Formula for Relative
  Entropy and its Asymptotics in Quantum Probability,
  \emph{Commun. Math. Phys.} 143, 99-114 (1991)

\bibitem{ogawa}
T. Ogawa, H. Nagaoka, Strong Converse and Stein's Lemma in Quantum Hypothesis 
Testing, \emph{IEEE Trans. Inform. Theo.}, vol. 46, No. 7, 2428-2433 (2000)



\bibitem{ohya}
M. Ohya, D. Petz, Quantum Entropy and its Use, Springer, Berlin 1993



\bibitem{ornstein}
D. Ornstein, B. Weiss, The Shannon-McMillan-Breiman Theorem
for a Class of Amenable Groups, \emph{Israel J. Math.} 44 (1), 53-60
 (1983)

\bibitem{ruelle} 
D. Ruelle, Statistical Mechanics, W.A. Benjamin, New York 1969

\bibitem{shields}
P. C. Shields, Two divergence-rate counterexamples, \emph{J. Theor. Prob.} vol. 6 521-545 (1993)
\end{thebibliography}
\end{document}